


\font\seventeenrm=cmr17

\font\twelverm=cmr12
\font\twelvei=cmmi12
\font\twelvesy=cmsy10 scaled 1200
\font\twelvebf=cmbx12
\font\twelvett=cmtt12
\font\twelveit=cmti12
\font\twelvesl=cmsl12
\font\twelvesc=cmcsc10 scaled 1200
\font\twelvess=cmss12
\font\twelvessi=cmssi12
\font\twelvebmit=cmmib10 scaled 1200
\font\twelvebsy=cmbsy10 scaled 1200

\font\elevenrm=cmr10 scaled 1095
\font\eleveni=cmmi10 scaled 1095
\font\elevensy=cmsy10 scaled 1095
\font\elevenbf=cmbx10 scaled 1095
\font\eleventt=cmtt10 scaled 1095
\font\elevenit=cmti10 scaled 1095
\font\elevensl=cmsl10 scaled 1095
\font\elevensc=cmcsc10 scaled 1095
\font\elevenss=cmss10 scaled 1095
\font\elevenssi=cmssi10 scaled 1095
\font\elevenbmit=cmmib10 scaled 1095
\font\elevenbsy=cmbsy10 scaled 1095

\font\tensc=cmcsc10
\font\tenss=cmss10
\font\tenssi=cmssi10
\font\tenbmit=cmmib10
\font\tenbsy=cmbsy10

\font\ninerm=cmr9    \font\eightrm=cmr8    \font\sixrm=cmr6
\font\ninei=cmmi9    \font\eighti=cmmi8    \font\sixi=cmmi6
\font\ninesy=cmsy9   \font\eightsy=cmsy8   \font\sixsy=cmsy6
\font\ninebf=cmbx9   \font\eightbf=cmbx8   \font\sixbf=cmbx6
\font\ninett=cmtt9   \font\eighttt=cmtt8
\font\nineit=cmti9   \font\eightit=cmti8   
\font\ninesl=cmsl9   \font\eightsl=cmsl8
\font\niness=cmss9   \font\eightss=cmss8
\font\ninessi=cmssi9 \font\eightssi=cmssi8


\skewchar\twelvei='177    \skewchar\eleveni='177     \skewchar\ninei='177
\skewchar\eighti='177     \skewchar\sixi='177
\skewchar\twelvesy='60    \skewchar\elevensy='60     \skewchar\ninesy='60
\skewchar\eightsy='60     \skewchar\sixi='60


\catcode`@=11
\newskip\ttglue
\parindent=3em

\def\twelvept{\def\rm{\fam0\twelverm}
   \textfont0=\twelverm \scriptfont0=\ninerm \scriptscriptfont0=\sevenrm%
   \textfont1=\twelvei  \scriptfont1=\ninei  \scriptscriptfont1=\seveni%
   \textfont2=\twelvesy \scriptfont2=\ninesy \scriptscriptfont2=\sevensy%
   \textfont3=\tenex    \scriptfont3=\tenex  \scriptscriptfont3=\tenex%
   \textfont\itfam=\twelveit   \def\it{\fam\itfam\twelveit}%
   \textfont\slfam=\twelvesl   \def\sl{\fam\slfam\twelvesl}%
   \textfont\ttfam=\twelvett   \def\tt{\fam\ttfam\twelvett}%
   \textfont\bffam=\twelvebf   \scriptfont\bffam=\ninebf%
      \scriptscriptfont\bffam=\sevenbf   \def\bf{\fam\bffam\twelvebf}%
   \def\oldstyle{\fam1 \twelvei}%
   \tt \ttglue=.5em plus.25em minus.15em%
   \normalbaselineskip=13pt plus.5pt minus.5pt%
   \def\doublespacing{\baselineskip=26pt plus.5pt minus 1pt}%
   \def\spaceandahalf{\baselineskip=19.5pt plus.5pt minus .5pt}%
   \setbox\strutbox=\hbox{\vrule height9pt depth4pt width0pt}%
   \let\sc=\twelvesc   \let\big=\tenbig%
   \let\ss=\twelvess   \let\ssi=\twelvessi%
   \let\bmit=\twelvebmit\let\bsy=\twelvebsy%
   \normalbaselines\rm}

\def\elevenpt{\def\rm{\fam0\elevenrm}
   \textfont0=\elevenrm \scriptfont0=\eightrm \scriptscriptfont0=\sixrm%
   \textfont1=\eleveni  \scriptfont1=\eighti  \scriptscriptfont1=\sixi%
   \textfont2=\elevensy \scriptfont2=\eightsy \scriptscriptfont2=\sixsy%
   \textfont3=\tenex    \scriptfont3=\tenex  \scriptscriptfont3=\tenex%
   \textfont\itfam=\elevenit   \def\it{\fam\itfam\elevenit}%
   \textfont\slfam=\elevensl   \def\sl{\fam\slfam\elevensl}%
   \textfont\ttfam=\eleventt   \def\tt{\fam\ttfam\eleventt}%
   \textfont\bffam=\elevenbf   \scriptfont\bffam=\eightbf%
      \scriptscriptfont\bffam=\sixbf   \def\bf{\fam\bffam\elevenbf}%
   \def\oldstyle{\fam1 \eleveni}%
   \tt \ttglue=.5em plus.25em minus.15em%
   \normalbaselineskip=12pt plus.5pt minus.5pt%
   \def\doublespacing{\baselineskip=24pt plus.5pt minus1pt}%
   \def\spaceandahalf{\baselineskip=18pt plus.5pt minus .5pt}%
   \setbox\strutbox=\hbox{\vrule height8.5pt depth3.5pt width0pt}%
   \let\sc=\elevensc   \let\big=\tenbig
   \let\ss=\elevenss   \let\ssi=\elevenssi%
   \let\bmit=\elevenbmit\let\bsy=\elevenbsy%
   \normalbaselines\rm}

\def\tenpt{\def\rm{\fam0\tenrm}
   \textfont0=\tenrm \scriptfont0=\sevenrm \scriptscriptfont0=\fiverm%
   \textfont1=\teni  \scriptfont1=\seveni  \scriptscriptfont1=\fivei%
   \textfont2=\tensy \scriptfont2=\sevensy \scriptscriptfont2=\fivesy%
   \textfont3=\tenex \scriptfont3=\tenex   \scriptscriptfont3=\tenex%
   \textfont\itfam=\tenit   \def\it{\fam\itfam\tenit}%
   \textfont\slfam=\tensl   \def\sl{\fam\slfam\tensl}%
   \textfont\ttfam=\tentt   \def\tt{\fam\ttfam\tentt}%
   \textfont\bffam=\tenbf   \scriptfont\bffam=\sevenbf%
      \scriptscriptfont\bffam=\fivebf   \def\bf{\fam\bffam\tenbf}%
   \def\oldstyle{\fam1 \teni}%
   \tt \ttglue=.5em plus.25em minus.15em%
   \normalbaselineskip=11pt plus.5pt minus.5pt%
   \def\doublespacing{\baselineskip=22pt plus.5pt minus 1pt}%
   \def\spaceandahalf{\baselineskip=16.5pt plus.5pt minus .5pt}%
   \setbox\strutbox=\hbox{\vrule height8.5pt depth3.5pt width0pt}%
   \let\sc=\tensc   \let\big=\tenbig%
   \let\ss=\tenss   \let\ssi=\tenssi%
   \let\bmit=\tenbmit\let\bsy=\tenbsy%
   \normalbaselines\rm}

\def\ninept{\def\rm{\fam0\ninerm}
   \textfont0=\ninerm \scriptfont0=\sixrm \scriptscriptfont0=\fiverm%
   \textfont1=\ninei  \scriptfont1=\sixi  \scriptscriptfont1=\fivei%
   \textfont2=\ninesy \scriptfont2=\sixsy \scriptscriptfont2=\fivesy%
   \textfont3=\tenex  \scriptfont3=\tenex \scriptscriptfont3=\tenex%
   \textfont\itfam=\nineit   \def\it{\fam\itfam\nineit}%
   \textfont\slfam=\ninesl   \def\sl{\fam\slfam\ninesl}%
   \textfont\ttfam=\ninett   \def\tt{\fam\ttfam\ninett}%
   \textfont\bffam=\ninebf   \scriptfont\bffam=\sixbf%
      \scriptscriptfont\bffam=\fivebf   \def\bf{\fam\bffam\ninebf}%
   \def\oldstyle{\fam1 \ninei}%
   \tt \ttglue=.5em plus.25em minus.15em%
   \normalbaselineskip=10pt plus.5pt minus.5pt%
   \def\doublespacing{\baselineskip=20pt plus.5pt minus1pt}%
   \def\spaceandahalf{\baselineskip=15pt plus.5pt minus .5pt}%
   \setbox\strutbox=\hbox{\vrule height8pt depth3pt width0pt}%
   \let\sc=\sevenrm   \let\big=\ninebig%
   \let\ss=\niness    \let\ssi=\ninessi%
   \normalbaselines\rm}

\def\eightpt{\def\rm{\fam0\eightrm}
   \textfont0=\eightrm \scriptfont0=\sixrm \scriptscriptfont0=\fiverm%
   \textfont1=\eighti  \scriptfont1=\sixi  \scriptscriptfont1=\fivei%
   \textfont2=\eightsy \scriptfont2=\sixsy \scriptscriptfont2=\fivesy%
   \textfont3=\tenex   \scriptfont3=\tenex \scriptscriptfont3=\tenex%
   \textfont\itfam=\eightit   \def\it{\fam\itfam\eightit}%
   \textfont\slfam=\eightsl   \def\sl{\fam\slfam\eightsl}%
   \textfont\ttfam=\eighttt   \def\tt{\fam\ttfam\eighttt}%
   \textfont\bffam=\eightbf   \scriptfont\bffam=\sixbf%
      \scriptscriptfont\bffam=\fivebf   \def\bf{\fam\bffam\eightbf}%
   \def\oldstyle{\fam1 \eighti}%
   \tt \ttglue=.5em plus.25em minus.15em%
   \normalbaselineskip=9pt plus.5pt minus.5pt%
   \def\doublespacing{\baselineskip=18pt plus.5pt minus1pt}%
   \def\spaceandahalf{\baselineskip=13.5pt plus.5pt minus .5pt}%
   \setbox\strutbox=\hbox{\vrule height7pt depth2pt width0pt}%
   \let\sc=\sixrm   \let\big=\eightbig%
   \let\ss=\eightss \let\ssi=\eightssi%
   \normalbaselines\rm}

\def\tenbig#1{{\hbox{$\left#1\vbox to8.5pt{}\right.\n@space$}}}
\def\ninebig#1{{\hbox{$\textfont0=\tenrm\textfont2=tensy
   \left#1\vbox to7.25pt{}\right.\n@space$}}}
\def\eightbig#1{{\hbox{$\textfont0=\ninerm\textfont2=ninesy
   \left#1\vbox to6.5pt{}\right.\n@space$}}}

\let\singlespacing=\normalbaselines



\def\today{\ifcase\month\or
   January\or February\or March\or April\or May\or June\or
   July\or August\or September\or October\or November\or December\fi
   \space\number\day, \number\year}

\def\sciday{\number\day
   \space\ifcase\month\or
   January\or February\or March\or April\or May\or June\or
   July\or August\or September\or October\or November\or December\fi
   \space\number\year}

\newcount\tyme
\newcount\hour
\newcount\minute

\def\amorpm{a.m.}
\def\tod{\gettime\number\hour:\ifnum\minute<10{}0\fi\number\minute\space\amorpm}

\def\gettime{\tyme=\time
   \divide \tyme by 60
   \hour=\tyme
   \ifnum\hour=12\gdef\amorpm{p.m.}\fi
   \ifnum\hour=0 \advance \hour by  12\fi
   \ifnum\hour>12\advance \hour by -12\gdef\amorpm{p.m.}\fi
   \multiply \tyme by 60
   \advance \time by -\tyme
   \minute=\time
   \advance \time by \tyme}



\def\underule#1{$\setbox0=\hbox{#1} \dp0=\dp\strutbox
    \m@th \underline{\box0}$}


\def\narrow{\advance\leftskip by3em \advance\rightskip by3em}
\def\wide{\advance\leftskip by-3em \advance\rightskip by-3em}


\def\alph#1{\ifcase#1\or a\or b\or c\or d\or e\or f\or g\or h\or i\or j\or
   k\or l\or m\or n\or o\or p\or q\or r\or s\or t\or u\or v\or w\or x\or
   y\or z\fi}

\def\deg{\ifmmode^\circ\else$^\circ$\fi}


\def\applt{\mathrel{\mathpalette\@versim<}}
\def\appgt{\mathrel{\mathpalette\@versim>}}
\def\@versim#1#2{\lower2pt\vbox{\baselineskip0pt \lineskip-.5pt
   \ialign{$\m@th#1\hfil##\hfil$\crcr#2\crcr\sim\crcr}}}


{\catcode`p=12 \catcode`t=12 \gdef\\#1pt{#1}}
\let\getfactor=\\
\def\kslant#1{\kern\expandafter\getfactor\the\fontdimen1#1\ht0}
\def\vector#1{\ifmmode\setbox0=\hbox{$#1$}%
    \setbox1=\hbox{\the\scriptscriptfont1\char'52}%
	\dimen@=-\wd1\advance\dimen@ by\wd0\divide\dimen@ by2%
    \rlap{\kslant{\the\textfont1}\kern\dimen@\raise\ht0\box1}#1\fi}

\tenpt\singlespacing

\vsize=9truein
\hsize=6.5truein
\twelvept
\rightline {KU-HEP-93-27}
\bigskip
\centerline{\seventeenrm Measuring Chirally Odd Wave Functions with Helicity
Flip Form Factors$^*$}
\vskip .5in

\centerline{{\bf Pankaj Jain} and {\bf John P. Ralston}}

\centerline{Department of Physics and Astronomy}

\centerline{University of Kansas, Lawrence, KS 66045}
\vskip .5in

\spaceandahalf
{\elevenpt
\centerline{\bf Abstract}
\medskip
We consider the role of chirally
odd wave functions in hard exclusive reactions.  Such wave functions have  the
quarks oriented in the opposite helicity configuration from those assumed in
the short-distance limit and  are generally associated with non-zero orbital
angular momentum. Calculations in the impulse  approximation allow for non-zero
helicity flip amplitudes while the conventional factorization prescription  for
exclusive processes does not.  By introducing a new approach, we show how
helicity flip form factors  are nevertheless calculable in QCD.}

\bigskip
\vfill
\noindent $^*$To appear in the proceedings of the workshop
of Future Directions in Particle and Nuclear Physics at Multi-GeV Hadron
Facilities, BNL, March 1993.
\eject
\noindent {\bf Factorization and Hadron Helicity Flip}
\medskip

The theory of elastic form factors at large momentum transfer in QCD is well
developed.   Nevertheless, conventional theory cannot be compared with data for
form factors involving hadronic  helicity flip.  This shortcoming is due to a
theoretical choice of ``factorization" scheme, and is not a  property of the
basic impulse approximations made in QCD.  The main difference between helicity
flip and non-flip quantities is that new wave functions appear in combinations
that are not reducible  to the quantities measured in non-flip reactions.  This
is a positive development: it means that the  internal structure of the quarks
in the hadron can be probed in a new way.
\medskip

In the conventional approach[1], the factorization of a typical form factor
$F(Q^2)$ reduces it to a  convolution over the momentum fractions $x_{i,j}$ of
the participating partons:
$$F(Q^2)=\int\prod_{ij} dx_idx_j\phi_j^*(x_j,\;
Q^2)H(x_i,x_j, Q^2)\phi_I(x_i, Q^2)\eqno (1)$$  Here, $H$ is a hard scattering
kernel, representing the part of the amplitude that is perturbatively
calculable; the $\phi$'s are objects called distribution amplitudes, which
contain
the non-perturbative  information about the hadrons.  This kind of
factorization prescribes a dynamical symmetry which  is manifested[1] in the
hadronic helicity conservation rule $\lambda_A+\lambda_B=\lambda_C+\lambda_D$
for reactions of the  form $A+B \to C+D$.  This rule is as general as the
factorization.  The key step is the
relation of the  distribution amplitudes $\phi$ to the wave functions $\psi$.
A useful
relation is obtained[2] in coordinate  space, letting b be the transverse
separation between a pair of quarks:
$$\phi(x,Q^2)=2\pi \int^\infty_0 db\; Q J_1(Qb)\tilde\psi_{m=0}(x,b)\eqno (2)$$
Here we have expanded the b-space wave function in orbital angular momentum
eigenstates with the  $z$-axis along the direction of the momentum; $J_1$ is a
Bessel function.  The wave function  participating in the model is {\it
selected} by
the hard scattering formalism to be carrying $m = 0$.  This  wave function is
called the ``short-distance" one.  By angular momentum conservation the quark
helicities for this case add up to be the hadron helicity.  The nearly perfect
chiral symmetry of  perturbative QCD predicts that quark spins do not flip in
the hard scattering, so that {\it in the model}  the sum of the hadron
helicities
cannot change.
\medskip

The success of the helicity conservation rule in comparison with data is
uneven.  There is a  consistent pattern of its violation in hadron-hadron
reactions.  For a long time this has been thought  to be a problem for
perturbative QCD.  Recently[2,3], it has been recognized that certain
independent scattering processes[4] disobey the presumed factorization (1),
even at large $Q^2$.  It has  been proposed[2] that these processes are likely
to
be the explanation for helicity violation in that  case.

\medskip

In certain photon initiated reactions the non-short distance aspects of
independent scattering seem  not to be a problem.   Thus (1) should apply at
leading order to form factors, and helicity non- conservation should be power
suppressed compared to helicity conservation. But being power  suppressed does
not say that the form factors are zero;  why does the factorization
prescription say  that they are zero?

\medskip A new approach[5] outlined here allows us to predict helicity flip
form factors, or, more objectively,  to interpret them as measurements of
hadron internal structure.  The cases in which we can do this  are those in
which the helicity flip term is the first, leading order term which is linearly
independent  and separable by its Lorentz structure.  For definiteness we
illustrate our study of the proton  magnetic form factor $F_2(Q^2)$.  We use
the impulse approximation and assume that we have perfect  chiral symmetry in
the hard collision.  Thus we have to understand how the quark, whose helicity
is  not flipped, can end up in a proton whose helicity is flipped. \medskip The
photon causes the scattering of a quark out of one proton and into the next
with momentum  transfer $Q^\mu$.  From crossing symmetry this can be related to
the antiquark-proton scattering with ($t$-channel) momentum transfer $Q$.  To
study this we introduce a new object: the off-diagonal, or  transition
amplitude we denote by $T$:  $$T^{ij}=\int d^4x\; e^{ikx} <p+Q,s\mid
T(\psi_i(0)\bar\psi_j(x)\mid p,s>\eqno (3)$$  Here $i, j$ are the Dirac indices
of
the quark field, which tell how its spin is oriented. In the diagonal  case
(same initial and final states, $Q=0$) the imaginary part of $T$ from the cut
between the quarks is  the parton distribution.  By definition $T$ has an
inclusive character: $T$ automatically sums  over all but  one Fock state
components, no matter what their momenta, spin, color, isospin, etc.
\medskip
The electromagnetic current is chirally invariant and thus only the
parts of $T$ satisfying $\lbrace T, \gamma^5\rbrace = 0$  can be probed with a
photon.  This
is what we mean by ``chirally even".  The opposite possibility is  to be
chirally odd, or $[T, \gamma^5] =0.$  The same selection rules occur in deeply
inelastic scattering,  where  certain inclusive parton distributions - the
unpolarized and the longitudinally polarized ones -  are chirally even and
``measurable", while the chirally odd transverse polarization distribution is
``unmeasurable" in that experiment[6].  The transverse polarization is a
leading order distribution  which can be measured with an anti-quark probe in
Drell-Yan lepton pair production.
\medskip
The helicity-flip form factors can
be investigated in terms of $T$, but one cannot go further to the level   of
factorization given in Eq.(1) and still save the calculation. The problem of
the helicity flip form  factor is the same problem as finding {\it a helicity
flip
amplitude in antiquark-proton scattering. }
\medskip
As mentioned earlier, it has been shown in $pp\to pp$ scattering that the
independent scattering  kinematics allows all possible orbital angular momenta
to participate in the scattering process.   When this occurs the hadron can
flip its helicity.  Is there a similar process in quark-proton  scattering
which can do the same?  An example independent scattering contribution is shown
in   figure 1.
\medskip
We believe that these types of processes are the unique configurations which
can do the job, except  for ``endpoint" contributions (which would totally
destroy the power counting and are apparently  Sudakov suppressed).  In general
we can decompose the Dirac structure of $T$:
$$\eqalign{T^{ij}=u_a(p,s)\bar u_b(p',s')t^{ij}_{ab}\cr
t^{ij}_{ab}=\Gamma^{ij}\cdot V\Gamma_{ab}\cdot U\cr}$$
where $\Gamma$ are matrices spanning the Dirac space (the set $1, \gamma^5,
\gamma^\mu,$ etc.), $V$ and $U$ are tensors
contracted with the $\Gamma$'s and functions of the invariants $Q^2$, $Q\cdot
k$, etc.  We have suppressed the flavor
dependence on the struck quark. The part of the tensor $t^{ij}_{ab}$ for the
helicity
flip of interest must be
be chirally even in indices $i, j$, since there is no quark helicity flip, and
odd in indices $a,b$ for the
proton helicity flip.  We can make a list of the invariant amplitudes forming
$t^{ij}_{ab}$ :

$$t^{ij}_ab = t_1 (\gamma^\mu)^{ij}(\gamma_\mu)_{ab}  + t_2 (\gamma^\mu)^{ij}
(i\sigma_{\mu\nu}(p-p')^\nu)_{ab} + t_3 (i\sigma_{\lambda\rho}(p-p')^\rho)^{ij}
(i\sigma^{\lambda\nu}(p-p')_\nu)_{ab} +...\eqno (4)$$
In (4) we indicate that there are many possible orthogonal Dirac projections,
which we do not  bother to write down.  What we want is the measurable ones,
which we now show are the first two.   Putting together the factors, the form
factor are calculated:
$$\eqalign{\bar u (p',s')(F_1\gamma^\mu+{iF_2\over 2m}\sigma^{\mu\nu}
q_\nu)u(p,s)&={1\over 4}\int d^4k\; Tr[\gamma^\mu T]\cr
&=\bar u (p',s')\gamma^\mu(ps)\int d^4k t_1 +\bar u(p',s')
i\sigma^{\mu\nu}q_\nu (ps)\int d^4kt_2\cr}\eqno (5)$$
The relation (5) is perfectly general, as can be checked by inserting the
definition (3) and obtaining  the matrix element of the electromagnetic current
operator $< p's'\mid J_{em}^\mu\mid ps>$.  We can read off the  form factors:
$$F_1=\int d^4kt_1;\hskip .5in F_2=\int d^4kt_2$$
Our physical picture is that quark hadron scattering measures non-zero orbital
angular momentum  in the wave function; with the insertion of a hard probe
which ``pinches the ends" of the scattering  shut, we obtain the form factors.
\medskip
Now consider power counting of the process.  From Fig. (1), we have two hard
gluons, two quark  wave functions contracted into distribution amplitudes, and
two wave functions for the relative  orbital angular momentum of the quarks.
Working in transverse separation space we will have a  Fourier transform
exp$(ib\cdot Q$) so that for large $Q$, each power of $b$ scales like $1/Q$.  A
wave function  carrying orbital angular momentum $m$ goes like $b^m$ as $b\to
0$.  We need
at least one power of $b$ for  a unit of orbital angular momentum, and then we
need two because the integration interval is  symmetric.  One then anticipates
the terms contributing to the form factor $F_2(Q^2)$ as:
$$F_2 (Q^2)=(1/Q^2)\alpha_s^2\prod_1 \int
dx_{i,j}d^2b\sum_{j,k}\phi_1^*\tilde\psi_J^* (b)e^{i Q\cdot
b}\phi_j\tilde\psi_k
(b)$$
where $j$ and $k$ are the indices in an orbital angular momentum expansion, and
we
let the $x$-dependence be implicit.  By power counting then, $F_2$ goes like
$1/Q^6$,
and is calculable if all the wave  functions are known.  It follows that a
measurement of $F_2$ measures the orbital angular momentum  content of the
quarks
in the proton.  Of course, a series of definitions is needed to say exactly
which wave functions are measured.  This is a new result - previously $F_2$ was
merely argued to be  ``higher twist" and the helicity flip was attributed to
quark mass terms.  Recent SLAC data[7] shows  that $F_2$ conforms to the power
counting above and is not anomalously small in magnitude.  This  allows us to
conclude that the projections onto $m=1$ wave functions of quarks in the proton
are  about the same size as the $m=0$ parts.  This is consistent with the
picture
that helicity flip in the  $pp\to pp$ hard scattering rate is not suppressed at
large $Q^2$, and occurs due to the intrusion of quark  orbital angular momentum
permitted by independent scattering.
\medskip
The spin structure of hadrons is growing more and more interesting.  The idea
that helicity flip  form factors measure non-zero orbital angular momentum can
be tested in color transparency:  quasiexclusive electroproduction knockout of
protons (or self-analyzing deltas) from a nuclear  target, with measured final
state polarization.  Non-zero $m$ should preferentially be filtered away at
large $A$.  We expect more theoretical work in this area, and exciting
interaction with experiment.
\bigskip

\noindent {\bf Acknowledgements}:  This work was supported in part under DOE
grant number 85-ER-FG02 40214. A008
\bigskip

\noindent {\bf References}
\medskip
\singlespacing
\item{1)} S. J Brodsky and G. P. LePage, {\it Phys. Rev.} {\bf D22}, 2157
(1980);
G. P. Lepage and S. J. Brodsky,  {\it Phys. Rev}. {\bf D22} , 2157 (1980).
\medskip
\item{2)} J. P. Ralston and B. Pire, ``High Energy Helicity Violation in Hard
Exclusive Scattering Hadrons"  Kansas and Ecole Polytechnique preprint
A175.0592, submitted to {\it Mod. Phys. Lett.}
\medskip
\item{3)} J. Botts and G. Sterman, {\it Nuc. Phys.} {\bf B325}, 62 (1989);  J.
Botts, ibid {\bf 353}, 20 (1989).
\medskip
\item{4)} P. V. Landshoff, {\it Phys Rev} {\bf D10}, 1024 (1974).
\medskip
\item{5)} P. Jain and J. P. Ralston, in preparation
\medskip
\item{6)} J. L. Cortes, B. Pire and J. P. Ralston, in {\it Polarized Collider
Workshop} (Penn State, 1990) edited by J. C. Collins {\it et al.} (APS \#223) p
176, and in {\it Zeit. Phys}. {\bf C55} 409 (1992); R.L Jaffe and X. Ji, {\it
Phys Rev Lett} {\bf 67}, 552, (1991).
\medskip
\item{7)} P. E. Bosted, {\it et al}, SLAC-PUB 5744 (1992).
\bye